\documentclass[12pt]{iopart}

\usepackage{pstricks}
\usepackage{graphicx}
\usepackage{dcolumn}
\usepackage{bm}

\usepackage{iopams}
\usepackage{pifont}
\usepackage{psfrag}
\usepackage[latin2]{inputenc}
\usepackage{cite}

\newcommand\ket[1]{|{#1}\rangle}

\newcommand\ibrkt[2]{|{#1}\rangle\langle{#2}|}

\newcommand{\be}{\begin{equation}}
\newcommand{\ee}{\end{equation}}

\newcommand{\ehoch}[1]{\rme^{#1}}
\newcommand{\skipc}[2]{}
\newcommand{\fig}[1]{Figure~\ref{#1}}
\newcommand{\eq}[1]{(\ref{#1})}

\begin{document}

\title[Factorization  with Gau{ss} sums: Implementations]{Factorization of numbers with Gau{ss} sums:\\
II. Suggestions for implementations with chirped laser pulses}

\author{W Merkel$^1$, S W\"olk$^1$,  W P Schleich$^1$, I Sh Averbukh$^2$, 
B Girard$^3$ and G G Paulus$^4$ }

\address{$^1$ Institut f\"ur Quantenphysik, Universit\"at Ulm,\\ Albert-Einstein-Allee 11, D-89081 Ulm, Germany 
}%

\address{$^2$ Department of Chemical Physics,\\ Weizmann Institute of Science, Rehovot 76100, Israel 
}

\address{$^3$ Laboratoire de Collisions, Agr\'egats, R\'eactivit\'e,   IRSAMC (Universit\'e de Toulouse/UPS; CNRS) Toulouse, France
}

\address{$^4$ Institut f\"ur Optik und Quantenelektronik, Friedrich-Schiller-Universit\"at Jena,\\ Max-Wien-Platz 1, D-07743 Jena, Germany 
}

\ead{sabine.woelk@uni-ulm.de}

\date{\today}

\date{\today}

\begin{abstract}
We propose three implementations of the Gauss sum  factorization schemes discussed in part I of this series [S W\"olk et al., preceding article]: 
({\it i}) a two-photon transition in a multi-level ladder system induced by a chirped laser pulse,
({\it ii}) a chirped one-photon transition in a two-level atom with a periodically modulated excited state, and 
({\it iii})  a linearly chirped one-photon transition driven by a sequence of ultrashort pulses.
For each of these quantum systems we show that the excitation probability amplitude is given by an appropriate Gauss sum. We provide rules how to encode the number $N$ to be factored in our system and how to identify the factors of $N$  in the fluorescence signal of the excited state.
\end{abstract}


\submitto{New J. of Physics}

\section{Introduction}
In \cite{merkel:grundlagen:2010} we have shown that Gauss sums are excellent tools to factor numbers. 
Throughout this article we have concentrated on the general principles and the underlying mathematical foundations. However, we have not addressed  physical implementations of Gauss sums.
The present article complements \cite{merkel:grundlagen:2010} and proposes three distinct realizations of  Gauss sum factorization.

Several experiments \cite{mehring:2007,mahesh:2007,gilowski:2008,bigourd:2008,weber:2008,peng:2008,tamma:2009,tamma:2009:b,sadgrove:2008,sadgrove:2009} have already successfully demonstrated factorization with the help of Gauss sums. However,   all of them have implemented the truncated Gauss sum
\be
{\cal A}_N^{(M)}(\ell) \equiv \frac{1}{M+1}\sum\limits_{m=0}^{M}\exp\left(-2\pi \rmi m^2 \frac{N}{\ell}\right),
\ee 
but none of the  Gauss sums introduced in \cite{merkel:grundlagen:2010}. Moreover, in all these experiments, except the recent one using the Michelson interferometer, the ratio $N/\ell$ had to be precalculated. In contrast
the three implementations of Gauss sum factorization proposed in the present paper, encode the number $N$ to be factored and the trial factor $\ell$  in two independent variables. As a consequence, the ratio $N/\ell$ does not  to be precalculated.

The first suggestion utilizes a two-photon transition in an equidistant ladder system driven by a chirped laser pulse. Since in this system the excitation probability amplitude is given by  the continuous Gauss sum
\be
S_N(\xi)\equiv \sum\limits_{m=-M}^{M}w_m \exp\left[2\pi \rmi \left(\pm m+\frac{m^2}{N}\right)\xi\right],
\ee
 discussed in \cite{merkel:PRA:part2}, it   can factor numbers.

In addition, we suggest  two alternative approaches  based on one-photon transitions in a laser-driven two-level system. 
Both  rely on quantum interference of multiple optical excitation paths, but lead to different quadratic phase factors. 

We consider a two-level system with a permanent dipole moment in the excited state. 
A cw-microwave field interacting with this dipole modulates the energy of the excited state and induces in this way an equidistant set of sidebands. The probability amplitude for a  one-photon transition caused by a chirped laser pulse is the sum over all possible excitation channels involving one optical photon as well as multiple quanta of the microwave field. The Gaussian nature of this sum arises from  the quadratic chirp of the laser pulse.

The second approach uses a multi-pulse excitation of a two-level system with a linearly chirped resonance
frequency. The probability amplitude of excitation is a sum of contributions arising from each  pulse. The quadratic phase dependence characteristic of a Gauss sum originates from the linear variation of the 
resonance frequency of the two-level system.

Throughout the article we neglect spontaneous emission since the interaction time with the pulses is much shorter than the decay time of the atomic level. This fact allows us to describe the system by the Schr\"odinger equation rather than a density matrix.

Our article is organized as follows: Since  we rely heavily on the physics of chirped laser pulses we first summarize in
Sec.~\ref{sec:essentials} the basic elements of this branch of optics.
In Sec.~\ref{factorization:twophoton} we then recall the main results of Ref.~\cite{merkel:PRA:part2} and, in particular, the expression for the excitation probability amplitude corresponding to a chirped two-photon transition in an harmonic ladder system. Starting from this formula we demonstrate  that this system allows us to factor numbers, and that in principle  a single realization of a factorization experiment can be employed to reveal the factors of another number.

Section \ref{sec:engineering} provides the basic elements of Secs. \ref{sec:phase} and \ref{sec:comb} by engineering a one-photon transition.
As a second physical system implementing Gauss sums we investigate in Sec.\ref{sec:phase} a two-level system with a permanent dipole moment driven by a microwave field. A chirped laser pulse interacts with the so-engineered Floquet ladder. The resulting probability amplitude is again given by a Gauss sum. 
Section \ref{sec:comb} is devoted to the analysis of the third system which features a linear variation of the resonance condition of a two-level atom exposed to a pulse train. This system provides  an alternative way to factor numbers based on Gauss sums. After a comparison of both  factorization schemes in Sec.~\ref{comparison} we conclude in Sec.~\ref{conclusions} by a summary of our results and a brief outlook.

\section{Chirped pulses: Essentials}
\label{sec:essentials}

Throughout the article we take advantage of the technology of chirped laser pulses. For this reason we briefly summarize in the present section the key ideas and formulas of this field.

A chirped laser pulse 
\be
E(t)\equiv{\cal E}_0\left[ \ehoch{-\rmi\omega_L t}\,f(t)+\textrm{c.c.}\right]
\label{eq:el:field}
\ee
consists of an amplitude ${\cal E}_0$, a carrier frequency $\omega_L$ and  a pulse shape function
\begin{equation}
\label{pulse}
f(t)\equiv f_0 \,
\exp\left[ -\frac{1}{2}(\Delta\omega f_0)^2 t^2\right].
\end{equation}
Here we have introduced the complex-valued amplitude
\be
f_0\equiv\sqrt{\frac{1+\rmi a}{1+a^2}},
\label{f0}
\ee
and the dimensionless parameter  
\be
a\equiv\Delta\omega^2 \phi''
\label{scaledchirp}
\ee
represents the second order dispersion. Moreover, $\Delta\omega$ denotes the bandwidth of the pulse and
$\phi''\equiv{\rmd^2} \phi(\omega)/{\rmd\omega^2}$
is a measure of the quadratic frequency dependence of the phase of the laser pulse.

When we substitute \eq{f0} into the exponent of the Gaussian in  \eq{pulse} we find that the pulse shape
\be
f(t)= f_0 \textrm{exp}\left( -\frac{\Delta \omega^2}{2(1+a^2)}t^2\right) \textrm{exp}\left( -\rmi\frac{a\Delta \omega^2}{2(1+a^2)}t^2\right).\label{eq:f(t)}
\ee
of a chirped pulse consists of the product of a real-valued Gaussian and a phase factor whose phase is quadratic in time.

Since the instantaneous frequency 
\be
\nu(t) \equiv \frac{\textrm{d}}{\textrm{d}t} \left( \frac{a\Delta \omega^2}{2(1+a^2)}t^2\right)  = \frac{a\Delta \omega^2}{1+a^2}t
\ee
of the pulse is the derivative of this phase with respect to time, the frequency changes linearly in time as the pulse switches on and off. For a positive value of $\phi''$ we find an increasing frequency, whereas  a negative $\phi''$ corresponds to a decreasing frequency.

\section{Chirping a two-photon transition}
\label{factorization:twophoton}

\begin{figure}
\centerline{\includegraphics[width=0.35\textwidth]{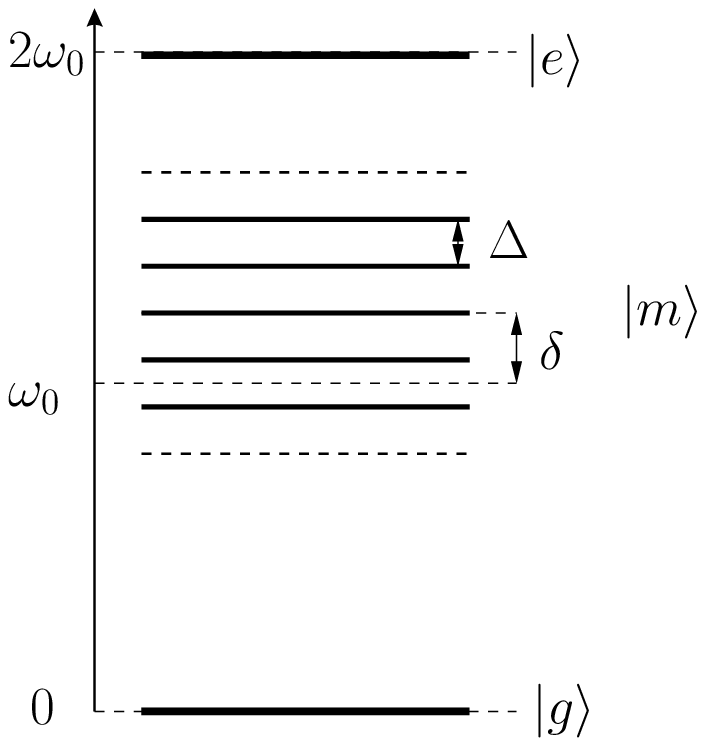}}
\caption{Model of ladder system. The ground state $\ket{g}$ is connected by a two-photon transition to the excited state $\ket{e}$. We include an harmonic manifold of $D\equiv M'+M+1$ intermediate states $\ket{m}$ with $M'\le m \le M$ which are shifted by the offset $\delta_m\equiv \delta+m\,\Delta$ with respect to the central frequency $\omega_0$. The offset of the central state with $m=0$ is $\delta$, whereas neighboring states in the harmonic manifold are separated by $\Delta$.}
\label{figure1a}
\end{figure}

In Ref. \cite{merkel:PRA:part2}  we have  considered a two-photon transition in the ladder system of \fig{figure1a} driven by a chirped laser pulse. In the weak field limit, the  probability amplitude to be in the excited state results from the interference of multiple quantum paths each contributing a quadratic phase factor. Since the population in the excited state has the form of a continuous Gauss sum  we can use this observable  to factor numbers as suggested\cite{merkel:grundlagen:2010} in part I of this series. 

We first briefly summarize the essential results of Ref.~\cite{merkel:PRA:part2}. Then we turn to a demonstration of the factorization capability and bring out the physical origin of the  scaling property derived in Ref.\cite{merkel:grundlagen:2010} from mathematical arguments.

\subsection{Brief review of the model}
We consider a quantum system with a ground state $\ket{g}$ and an excited state $\ket{e}$ separated by an energy $2\hbar \omega_0$. In the neighborhood of the midpoint of the  energy difference we assume to have a manifold of equidistant energy levels as shown in \fig{figure1a}. Their off-set 
\be
\delta_m \equiv \delta + m \Delta
\ee
with respect to the central frequency $\omega_0$ is the sum of the off-set $\delta$ to the central state and integer multiples $m$ of the separation $\Delta$ of two neighboring states of the manifold. 

At this point it is important to note that this decomposition of $\delta_m$ is not unique. Indeed, we could have chosen a "central" level which is different from the one indicated in \fig{figure1a}. This choice would have changed the integers $m$. This ambiguity in the labeling of the states is the deeper physical origin of the scaling property of the Gauss sum already mentioned in Ref.\cite{merkel:grundlagen:2010}.

Next we drive this ladder system by a chirped laser pulse of the form given by \eq{eq:el:field}.
In second order perturbation theory  the  probability amplitude  
\be
c_e^{(TPT)} = e^{\rmi\gamma}S_N \label{eq:ampli}
\ee
to be in the excited state after such a two-photon transition is given\cite{merkel:PRA:part2},
apart from the phase factor $\exp(i \gamma)$ by the Gauss sum
\begin{equation}
S_N(\xi)\equiv
\sum\limits_{m=-M'}^M w_m \,
\exp\left[2\pi \rmi\left(m+\frac{m^2}{N}\right)\xi\right]
\label{truncated:gauss}
\end{equation}
discussed in part I of this series. Here the variable
\begin{equation}
\xi\equiv\frac{\delta\Delta}{\pi} \phi''
\label{argument}
\end{equation}
is expressed in terms of the parameters $\delta$ and $\Delta$ of the harmonic manifold of intermediate states and the dimensionless chirp $\phi''$. Hence, by varying the chirp we can tune $\xi$. For this reason we call the variable $\xi$ the dimensionless chirp.

The number

\begin{equation}
N\equiv\frac{2\delta}{\Delta}
\label{number}
\end{equation}
to be factored is represented by the ratio of the two characteristic frequencies of the ladder.

The weight factors 
\be
w_m\equiv \tilde{\omega}_m
\textrm{erfc}\left(\rmi\frac{\delta_m}{\Delta\omega}\sqrt{1-\rmi a}\right)
\exp{\left[-\left(\frac{\delta_m}{\Delta\omega}\right)^2\right]}
\label{eq:weight}
\ee
contain the complementary error function\cite{abramowitz}
\be
\textrm{erfc}(z)\equiv \frac{2}{\sqrt{\pi}}\int\limits_z^{\infty}\rmd u\; \ehoch{-u^2}
\ee
of complex argument $z$, and the abbreviation
\be \tilde{\omega}_m \equiv
-\frac{\pi}{2} \frac{\Omega_{em}\Omega_{mg}}{\Delta\omega^2} 
\ee
involves the Rabi frequencies $\Omega_{mg}$ and $\Omega_{em}$ connecting the ground and the excited state with the intermediate states, respectively.

\subsection{Factorization}
\label{factorization:example}

We are now in the position to discuss  our factorization scheme. For this purpose we first address the experimental requirements and  limitations and then present an example demonstrating the capability of this system to factor numbers.

\subsubsection{Requirements}

The probability amplitude $c_e^{(TPT)}$, \eq{eq:ampli}, of populating the excited state is of the form of a continuous Gau{ss} sum discussed in part I of this series. Thus the dependence of the population $|c_e^{(TPT)}|^2$ in the excited state on  the dimensionless chirp $\xi$ can be employed to reveal the factors of an integer $N$.  For this purpose we encode $N$ according to \eq{number}  in the parameters $\delta$ and $\Delta$ of the harmonic manifold of intermediate states. In order to apply the factorization scheme to a broad range of numbers $N$, we require control over $\delta$ and $\Delta$.

We emphasize that the equidistant spacing within the harmonic manifold is essential for obtaining the Gauss sum and for our factorization scheme. Moreover, the  dimension $D\equiv M'+M+1$ of the intermediate levels has to be adapted to the number $N$ to be factored. The larger $N$ the more intermediate states are required for a meaningful signal. For the factorization of $N$ we require  $N\leq D \leq 2N$  where the lowest quantum number is bound by $M'>N/2$.

We can read out the population of the excited state by its fluorescence.
In Ref.\cite{merkel:grundlagen:2010} we have formulated a rule for determining the factors of $N$: If the signal shows a distinct  maximum around the integer  $\xi=\ell$ then $\ell$ is a factor of $N$. In order to resolve the signal in the vicinity of candidate prime numbers, we require sufficient stability of and accuracy in the chirp $\phi''$.

According to Ref.\cite{merkel:grundlagen:2010} we also need to impose a restriction on the weight factor $w_m$, \eq{eq:weight} of the contribution  arising from the quantum path through the $m$th intermediate state: Indeed, $w_m$ must  be slowly varying as a function of  $m$ in order to ensure that no specific excitation path is favored or discriminated. As a result the dipole matrix elements associated with all possible transitions should be of the same order of magnitude.

\subsubsection{Example}

Next we present numerical results  for an artificial ladder system consisting of $D$ intermediate states. 
Here, we have made the idealized assumption that the Rabi frequencies associated with the sequential path $\ket g\to \ket m \to \ket e$ are identical, that is $\Omega_{em}\Omega_{mg}=\textrm{const.}$.

\begin{figure}[ht]
\begin{center}
\includegraphics[width=0.65\textwidth]{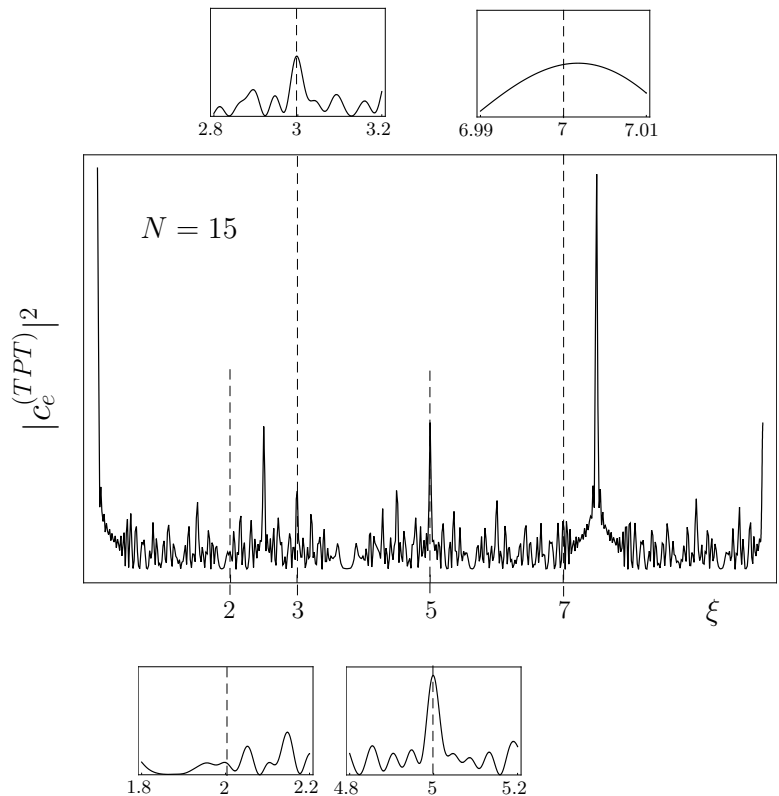}
\end{center}
\caption{Factorization of $N=15=3\cdot 5$ with the help of the  population $|c_e^{(TPT)}|^2$ of the excited state given by \eq{eq:ampli} and \eq{truncated:gauss} after a chirped two-photon transition through the intermediate levels of the ladder system of \fig{figure1a}.
In the center we provide an overview over the complete signal as a function of the dimensionless chirp $\xi$. The insets magnify the signal in the vicinity of candidate prime factors. Pronounced maxima at the prime factors $\xi=3 \textrm{ and }5$ are clearly visible. In contrast, at non-factors $\xi=2 \textrm{ and }7$ the signal does not exhibit any peculiarities. Parameters are $\delta= 0.0225\textrm{fs}^{-1}$, $\triangle= 0.003\textrm{fs}^{-1}$, $\triangle w=  0.1525\textrm{fs}^{-1}$, $a=  -10824$ and $2M+1=23$.
}
\label{figure3}
\end{figure}

In \fig{figure3} we display the population $|c_e^{(TPT)}|^2$ of the excited state  for the number $N=15=3\cdot 5$  as a function of the dimensionless chirp $\xi$. For a single intermediate state, there exist\cite{merkel:PRA:part2} several interfering quantum paths only if $\xi < 0$. However, for an equidistant manifold, the population is symmetric with respect to $\xi$, that is $|c_e^{(TPT)}(\xi)|^2= |c_e^{(TPT)}(-\xi)|^2$.

The insets magnify the signal in the vicinity of trial factors $\xi=2,\,3,\,5 \textrm{ and } 7$. Following our criterion of a dominant maximum at an integer indicating a factor we clearly identify from the  distinct maxima  at $\xi=3\textrm{ and }5$ the factors of $N=15$. In contrast, the signal at non-factors such as $\xi=2$ or $7$ does not show any characteristic features.

\subsection{Factorization by rescaling}
\label{factorization:rescaling}
In part I of this series \cite{merkel:grundlagen:2010} we have already shown that the realization of the Gauss sum $S_N$ for $N$ is sufficient to reveal the factors of another number $N'$. Whereas the proof presented in Ref.\cite{merkel:grundlagen:2010} had relied on mathematical arguments we now use the freedom in labeling the levels of the equidistant ladder system to verify this surprising feature.

For this purpose we recall from \eq{number}  that  $N$ 
is encoded in the ratio $2 \delta / \Delta$ described by the offset $\delta$ of the reference level $\ket{m=0}$ and the level spacing $\Delta$. However, when we choose a different reference level $\ket{k}$ with the associated offset
\begin{equation}
\delta'\equiv \delta+k \Delta,
\end{equation}
the number to be factored is
\begin{equation}
N'\equiv\frac{2\delta'}{\Delta}=N+2k.
\end{equation}
With the help of the definition  \eq{number}
for the number $N$ to be factored   in terms of the offset $\delta$ and the detuning $\Delta$ we find from \eq{argument} that the  dimensionless chirp
\be
\xi=N\frac{\Delta^2}{\pi}\phi''
\ee
is proportional to $N$.

As a result, the dimensionless chirp
\be
\xi'\equiv N'\frac{\Delta^2}{\pi}\phi''
\ee
corresponding to $N'$ is related to $\xi$ and $N$ by the scaling transformation
\be
\xi' = \frac{N'}{N} \xi.
\ee
Therefore, we can factor the number $N'$ by analyzing the signal $c_e^{(TPT)}$, which was recorded in its dependence on $\xi$ for $N$, using the new scale $\xi'\equiv (N'/N)\cdot \xi$.

\section{Engineering a one-photon transition}
\label{sec:engineering}

In the preceding section we have used a chirped two-photon transition going through an equidistant ladder system to factor numbers. Unfortunately, the requirements on the system are rather stringent and it is hard to identify quantum systems with such an arrangement of levels. For these reasons we now study more elementary models based on a two-level system with a ground state $\ket{g}$ and an excited state $\ket{e}$ separated by an energy $\hbar\omega_0$. 

We assume the excited state to have a permanent dipole moment $\wp_{ee}$ which interacts with a time-dependent modulating field $E_m=E_\textrm{m}(t)$.  In the following sections we consider two cases: ({\it i}) a sinusoidal time dependence manifesting itself in a periodic modulation of the excited state, and ({\it ii}) a quadratic chirp reflecting itself in a linear shift. 

In addition to $E_\textrm{m}$ we have a  time-dependent weak driving field $E_\textrm{d} = E_\textrm{d}(t)$ causing transitions between the ground and the excited state. Depending on the two cases $E_\textrm{d}$ is either a single chirped laser pulse, or a sequence of pulses.

This arrangement corresponds to  the interaction Hamiltonian
\be
V\equiv-
\wp_{ee}E_\textrm{m}(t)\ibrkt{e}{e}
-\left(\wp_{ge}\ibrkt{e}{g}
+\textrm{c.c.}\right)
E_\textrm{d}(t)
\label{interaction1}
\ee
where $\wp_{ge}$ denotes the dipole moment of the two-level transition. Here we have assumed that the frequencies of $E_\textrm{m}$ and $E_\textrm{d}$  are clearly separated. Hence, $E_\textrm{m}$ only acts on the excited state and  $E_\textrm{d}$ only on the transition.

In the interaction picture the equations of motion for the probability amplitudes 
$c_e = c_e(t)$ and $c_g = c_g(t)$ to be in the excited and ground state read\cite{allen}

\be
\rmi\frac{d}{dt}\,c_e(t)=-\Omega_{ee}(t)\,c_e(t)-\Omega_{ge}(t)\ehoch{\rmi\,\omega_0 t}\,c_g(t)\label{eq:c_e(t)}
\ee
or
\be
\rmi\frac{d}{dt}\,c_g(t)=-\Omega_{eg}(t)\ehoch{-\rmi\,\omega_0 t}\,c_e(t)\label{eq:c_g(t)}.\phantom{-\Omega_{ee}(t)c_e(t)}
\ee

Here  we have defined the {\it time-dependent} Rabi-frequencies
\be
\Omega_{ee}(t) \equiv \wp_{ee} \,E_\textrm{m}(t)/\hbar
\quad \textrm{and}\quad
\Omega_{eg}(t) \equiv \wp_{eg} \,E_\textrm{d}(t)/\hbar
\ee
associated with the two electric fields $E_\textrm{m}$ and $E_\textrm{d}$, respectively.

In order to solve \eq{eq:c_e(t)} and \eq{eq:c_g(t)} we first recall that the strong field $E_\textrm{m}$ which causes the modulation of the excited state appears through $\Omega_{ee}$ in \eq{interaction1} and multiplies $c_e$. Only the weak driving field  $E_\textrm{d}$ which enters \eq{interaction1} via $\Omega_{ge}$ and multiplies $c_g$ induces transitions. For this reason it suffices to describe this process by perturbation theory of first order.

At time $t_0$ the two-level system occupies the ground level, that is $c_g(t_0)=1\quad\textrm{and}\quad c_e(t_0)=0$. In the weak field limit  the  probability amplitude for the excited state does not change significantly under the action of a weak chirped pulse which yields
$c_g(t)\approx 1$. Hence, \eq{eq:c_e(t)} reduces to the inhomogeneous differential equation
\begin{equation}
\rmi\frac{d}{dt}\,c_e(t)\cong
-\Omega_{ee}(t)c_e(t)-\Omega_{ge}(t)\ehoch{\rmi\,\omega_0t}
\label{firstorder:deq}
\end{equation}
where the interaction with the chirped laser pulse acts as an inhomogeneity.

It is easy to verify that the  solution of \eq{firstorder:deq} reads
\be
c_e(t)=\rmi\,e^{\rmi\, \beta(t)}
\int\limits_{t_0}^t dt'\, \exp[-\rmi\, \beta(t')]\exp(\rmi\,\omega_0 t') \,\Omega_{ge}(t') \,
\label{formal:solution}
\ee
where we have introduced 
the phase
\be
\beta(t)\equiv \int\limits_{t_0}^t dt'\; \Omega_{ee}(t')= \frac{\wp_{ee}}{\hbar}\int\limits_{t_0}^{t}dt'\; E_m(t').
\label{beta}
\ee

When we substitute the electric field 
\be
E_\textrm{d}(t)\equiv {\cal E}_d\left[ e^{-\rmi\omega_L t }h(t) + \textrm{c.c.}\right] \label{eq:Ed}
\ee
of amplitude ${\cal E}_d$, carrier frequency $\omega_L$ and envelope $h=h(t)$
 into the solution \eq{formal:solution} for the probability amplitude $c_e$ we find in rotating wave approximation
\be
c_e(t)=\rmi\,\Omega_{ge}\, e^{\rmi\, \beta(t)}
\int\limits_{t_0}^t dt'\, \exp[-\rmi\, \beta(t')]\exp(i\delta t') \,h(t').
\label{master:solution}
\ee
Here we have defined the {\it time-independent} Rabi frequency
\be
\Omega_{ge} \equiv \wp_{ge}\,{\cal E}_d/\hbar
\ee
associated with the electric field ${\cal E}_d$ of the transfer pulse and the detuning $\delta\equiv\omega_0-\omega_L$ between
the atomic and the carrier frequency.

\section{Floquet ladder}
\label{sec:phase}
\begin{figure}
\centerline{
\includegraphics[width=0.65\columnwidth]{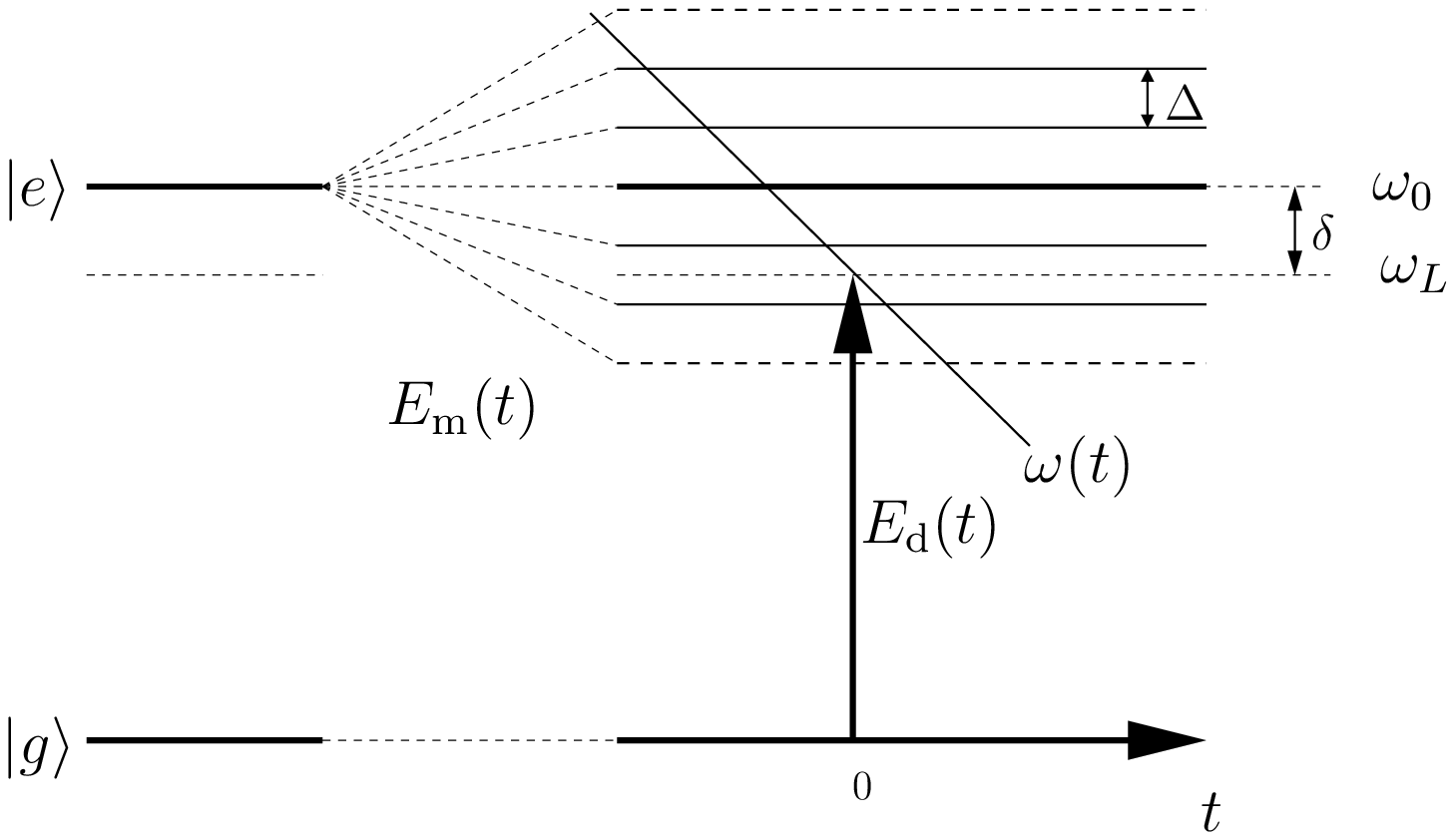}
}
\caption{Engineering the Floquet ladder.
We consider a two-level system with the ground state $\ket{g}$ and the excited state $\ket{e}$ separated by the energy $\hbar\omega_0$.
The excited state is modulated by a strong sinusoidal field $E_\textrm{m}=E_\textrm{m}(t)$, \eq{driving:cw} giving rise to equidistant sidebands separated by $\hbar\Delta$. The one-photon transition is driven by a chirped laser pulse $E_\textrm{d}=E_\textrm{d}(t)$, \eq{eq:Ed},  characterized by a linear variation of the instantaneous frequency $\omega=\omega(t)$.}
\label{figure1}
\end{figure}

So far we have neither specified the modulating nor the driving field. In the present section we consider a sinusoidal
modulation of the excited state by a strong cw field creating a set of equidistant sidebands,  very much in the spirit of the harmonic manifold of Figure \ref{figure1a}. Moreover, we include a weak chirped laser pulse driving the transition, that is the envelope $h=h(t)$ of $E_d$ is given by $f(t)$ of \eq{pulse}. Figure \ref{figure1} summarizes this engineering of the Floquet ladder.

\subsection{Excitation probability in the weak field limit}
We now evaluate the probability amplitude $c_e^{(FL)}$ given by \eq{beta} and \eq{master:solution} for the  excitation of the Floquet ladder. Here we proceed in two steps: we first include the modulation and then calculate the remaining integral for the case of a chirped pulse.
\subsubsection{Sinusoidal modulation}
In the case of  the modulation field
\be
E_\textrm{m}(t)\equiv{\cal F}_0 \cos (\Delta t+\varphi),
\label{driving:cw}
\ee
with period $2\pi/\Delta$, amplitude ${\cal F}_0$ and phase  $\varphi$,
the time-dependent phase $\beta=\beta(t)$ defined by \eq{beta} takes the form
\be
\beta(t)\equiv \kappa \sin(\Delta t+\varphi).
\ee
Here we have chosen the lower integration limit $t_0\equiv (n_0\pi-\varphi)/\Delta$ and have introduced the dimensionless ratio
\be
\kappa \equiv \frac{\Omega_{ee}}{\Delta}
\label{kappa}
\ee
of  the {\it time-independent} Rabi frequency
$
\Omega_{ee} \equiv \wp_{ee} \;{\cal F}_0/\hbar
$
and $\Delta$.

When we apply the generating function\cite{abramowitz} 
\be
\exp\left(\rmi\,\kappa \sin\theta \right)= 
\sum\limits_{n=-\infty}^\infty J_n(\kappa)\, e^{\rmi\,n\theta} 
\ee
of the Bessel function $J_n$ to evaluate the phase factor in the integrand of \eq{master:solution} we find   the probability amplitude
\be
c_e^{(FL)}(t)=\rmi\,\Omega_{ge} \, \rme^{\rmi\,\beta(t)}
\sum\limits_n J_n(\kappa)\,\ehoch{-\rmi\,n\varphi}
h_n(t)\label{eq:c_e(t)2}
\ee
to be in the excited state. 
Here we  have interchanged the order of integration and summation and have introduced the integral 
\be
h_n(t)\equiv \int\limits_{t_0}^t \rmd t' 
\exp\left(\rmi\,\delta_n t'\right)\,h(t')\label{eq:h(t)}
\ee
with the offset
\be
\delta_n\equiv \delta-n\Delta
\label{quasi:en}
\ee
of the $n$-th satellite of the excited state in the splitted manifold.

Hence, the  modulation  of the excited state causes equidistant sidebands  and  $J_n(\kappa)$ determines the weight of the $n$-th sideband. In the language of Floquet theory we have replaced the time dependent Hamiltonian by an infinite dimensional Floquet matrix.

\subsubsection{Chirped pulse}
So far our calculation is valid for an arbitrary pulse shape $h=h(t)$. 
We now perform the integration for the case of a chirped pulse where the envelope $h$ is given by the complex-valued Gaussian $f$ defined by \eq{pulse}.

Since we are interested in times after the pulse has interacted, that is for 
$ \sqrt{1+a^2}/\Delta\omega \ll t$ we extend the upper and lower limits of the integration in \eq{eq:h(t)} to $+\infty$ and $-\infty$, respectively, which yields 
\be
h_n(t) \cong h_n\equiv
f_0\int\limits_{-\infty}^\infty \rmd t\; \ehoch{-\frac{1}{2}(\Delta\omega\,f_0 t)^2+\rmi\,\delta_n t},
\ee
or
\be
h_n=\frac{\sqrt{2\pi}}{\Delta\omega} \exp\left[-\frac{1}{2f_0^2}\left(\frac{\delta_n}{\Delta\omega}\right)^2\right].\ee
When we recall the definition \eq{f0} of $f_0$ we can decompose this Gaussian into a real-valued one and into a quadratic phase factor, that is 
\be
h_n=\frac{\sqrt{2\pi}}{\Delta\omega}
\exp\left[-\frac{1}{2}\left(\frac{\delta_n}{\Delta\omega}\right)^2\right]
\exp\left[\rmi\,\delta_n^2 \frac{\phi''}{2}\right].\label{eq:hschlange2}
\ee
Here we have also made use of \eq{scaledchirp}.

\subsection{Emergence of Gauss sum}

In the expression for $h_n$ given by \eq{eq:hschlange2}  the off-set $\delta_n$ which  depends linearly on $n$ enters quadratically. Hence, $h_n$ contains quadratic phase factors. In the present section we cast the probability amplitude $c_e^{(FL)}$ given by \eq{eq:c_e(t)2}  into the form of a Gauss sum which allows  us to factor numbers.

With the help of \eq{eq:hschlange2} we find from \eq{eq:c_e(t)2} the formula

\be \fl
c_e^{(FL)}=\rmi\Omega_{ge}\textrm{e}^{\rmi\beta(t)}\frac{\sqrt{2\pi}}{\Delta w}\sum\limits_n \exp\left[-\frac{1}{2}\left(\frac{\delta_n}{\Delta w}\right)^2\right] J_n(\kappa)e^{-\rmi n\varphi}
\exp\left[\rmi\,\delta_n^2\, \frac{\phi''}{2}\right]\label{eq:38}.
\ee

Next we recall the definition \eq{quasi:en} of $\delta_n$ and express it by
\be
\frac{\delta_n}{\Delta w} = -\frac{n-\delta/\Delta}{\Delta w/\Delta} \equiv -\frac{n-N}{\Delta n},\label{eq:delta/w}
\ee
where in the last step we have introduced  the abbreviations
\be
N\equiv\frac{\delta}{\Delta}\quad \textrm{and}\quad \Delta n \equiv \frac{\Delta w}{\Delta}\label{def:N}.
\ee

Likewise, we obtain from \eq{quasi:en} the identity
\be
\delta_n^2=\delta^2- (n-n^2\frac{\Delta}{2\delta})2\delta\Delta.
\label{eq:42}
\ee

When we substitute \eq{eq:delta/w} and \eq{eq:42} into \eq{eq:38}, define the dimensionless chirp
\be
\xi \equiv \frac{\delta\Delta}{\pi}\phi''
\ee
and recall the definition \eq{def:N} of $N$
 we arrive at the probability amplitude 
\be
c_e^{(FL)}(t)={\cal N}(t)\sum\limits_n \tilde w_n
\exp\left[-\rmi\pi \left(n-\frac{n^2}{2N}\right)\xi\right]\label{eq:c_efinal},
\ee
to be in the excited state in the Floquet-ladder scheme.
Here, we have used the abbreviation
\be
{\cal N}(t)\equiv 2\pi \rmi \frac{\Omega_{ge}}{\Delta}\rme^{\rmi\beta(t)}\rme^{\rmi\delta^2\phi''/2}
\ee
together with the weight factor
\be
\tilde w_n \equiv \frac{1}{\sqrt{2\pi} \Delta n}\exp\left[-\frac{1}{2}\left(\frac{n-N}{\Delta n}\right)^2\right]J_n(\kappa) \rme^{-\rmi n\varphi}\label{def:w_n}.
\ee

In order to reduce the influence of the Bessel function $J_n$ in  $\tilde w_n$ we adjust the modulation index
\be
\kappa\equiv \frac{\Omega_{ee}}{\Delta}=\frac{{\cal F}_0 \wp_{ee}}{\hbar \Delta}
\ee
determined by the microwave field, \eq{driving:cw}, such that  $\tilde w_n$ is slowly varying as a function of $n$.
For this purpose we recall\cite{abramowitz} the asymptotic expansion 
\be
J_n(z)\stackrel{\sim}{=}
\sqrt{\frac{2}{\pi z}}\cos\left(z-n\,\frac{\pi}{2}-\frac{\pi}{4}\right)
\ee
of the Bessel function in the limit of large arguments $n \ll z$.
Indeed, we find for the choice 
\be
\kappa\equiv 2 \pi s +\frac{\pi}{4},
\ee
where $s$ is a large integer the approximation

\be
J_n(\kappa)
\stackrel{\sim}{=}
\sqrt{\frac{2}{\pi \kappa}}\,
\left\{\begin{array}{cl}
(-1)^m &\textrm{for } n=2m\\[2mm]
0      &\textrm{for } n=2m+1
\end{array}\right..
\label{weightfactor}
\ee
Thus all weight factors $\tilde w_n$ with odd index $n$ vanish and the probability amplitude $c_e^{(FL)}$ given by \eq{eq:c_efinal} reduces to
\be
c_e^{(FL)} \cong {\cal N} \sum\limits_m w_m \exp\left[-2\pi \rmi (m-\frac{m^2}{N})\xi\right], \label{peg2}
\ee
where
\be
w_m \equiv \frac{1}{\pi\Delta n} \frac{1}{\sqrt{\kappa}}\exp\left[-2\left(\frac{m-N/2}{\Delta n}\right)^2\right]e^{\rmi m(\pi -2\varphi)}.\label{weight}
\ee

We note that for the choice of $|\varphi|=\pi/2$  the phase factor in $w_m$ is unity and the only $m$-dependence left  results from the Gaussian. For an appropriate choice of $\Delta n$, which according to \eq{def:N} is determined by the bandwidth  $\Delta\omega$ of the chirped pulse, this Gaussian is slowly varying.

When we compare the probability amplitude $c_e^{(FL)}$ to be in the excited state given by \eq{peg2}  to the generic representation
\be
{\cal S}(\xi;A,B)\equiv \sum_m w_m\, \exp\left[2\pi \rmi\,\left(\frac{m}{A}+\frac{m^2}{B}\right)\xi\right]\\
\label{peg}
\ee
of a Gauss sum discussed in part I of this series\cite{merkel:grundlagen:2010}  we find
\be
c_e^{(FL)}= {\cal N} {\cal S}(\xi;-1,N)\label{eq:c_e=S}.
\ee
Since the fluorescence signal of the excited state is proportional to the population $|c_e^{(FL)}|^2$ in this state, it is proportional to the Gauss sum $|{\cal S}(\xi;-1,N)|^2$.

\subsection{Factorization}
\label{factor1}

In order to  gain information on the factors of an appropriately  encoded number $N$ we present now two schemes: The first one requires a continuous measurement of the fluorescence signal as a function of the dimensionless chirp $\xi$. For the second one  it suffices to acquire the fluorescence signal at integer values $\xi=\ell$. 

\subsubsection{Continuous tuning of chirp}
\label{factor1:continuous}
\begin{figure}
\begin{center}
\includegraphics[width=0.5\columnwidth]{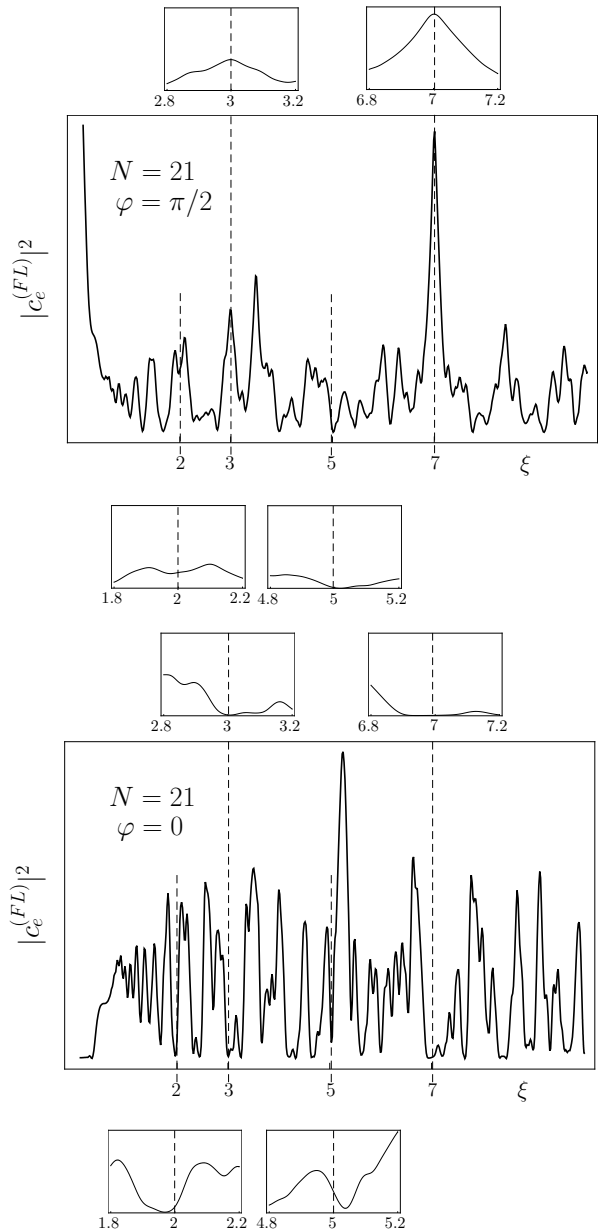}
\end{center}
\caption{Factorization of $N=21=3\cdot 7$ with $78$ satellites in the Floquet ladder of the excited state. 
Here we show the fluorescence signal $|c_e^{(FL)}|^2$, \eq{eq:c_efinal}, as a function of the continuous rescaled chirp $\xi$ for the phase $|\varphi|=\pi/2$ (top) and $\varphi = 0$ (bottom) of the cw-field. The electric field parameters are chosen to yield  the width of the weight factor distribution   $\Delta n=12.71$ and the modulation index $\kappa=100 \cdot  2\pi+\pi/4$.
The positions of candidate prime factors are indicated by vertical lines. The insets demonstrate that the signal exhibits pronounced maxima at the prime factors $\ell=3$ and $7$ (top) but not at $\ell=2$ and $5$ (bottom).
For the phase $\varphi=0$ the factorization criterion of observing distinct maxima at factors of $N$ does not apply. Here the factors are identified by zeros (top) rather than maxima. The non-factors have a non-vanishing signal (bottom).
}
\label{figure21}
\end{figure}

According to Ref.\cite{merkel:grundlagen:2010} $\ell$ is a factor, or a multiple of a factor, of $N$ if ${\cal S}(\xi;-1,N)$ given by \eq{peg} shows a pronounced maxima at $\xi=\ell$.
In \fig{figure21} we present numerical results for the factorization of $N=21=3\cdot 7$ employing a truncated Floquet ladder covering  $2M+1=78$ harmonics for two choices of the relative phase $\varphi$. Here, we display $|c_e^{(FL)}|^2$ based on \eq{eq:c_efinal}. Again we indicate candidate prime factors by vertical lines. 

For $\varphi=\pi/2$ we clearly identify from the insets on the top the prime factors  $\ell=3$ and $7$. In contrast, the signal does not show any peculiarities at non-factors such as $\ell=2$ and $5$ as shown by the insets at the bottom.

For $\varphi=0$ the phase factor $e^{\rmi m\pi}=(-1)^m$ leads to oscillatory weight factors and our criterion of finding pronounced maxima at factors of $N$ is not applicable here. Nevertheless, the signal $|c_e^{(FL)}|^2$ still contains information on the factors of $N$. Indeed, the corresponding signal shown at the bottom of \fig{figure21} vanishes at the factors $\ell=3 \textrm{ and }7$ as indicated by the insets on the top but displays no peculiarities at non-factors such as $\ell=2$ and $\ell=5$ depicted at the bottom. 

\subsubsection{Discrete values of chirp}
\label{factor1:discrete}
Next we present another approach towards factorization with the help of the Gauss sum, \eq{peg}. For this technique we assume that we have sufficient control over the rescaled chirp $\xi$ to tune it precisely to an integer $\xi=\ell$.
As a consequence, the term linear in the summation index in the phase factor drops out and the probability amplitude $c_e^{(FL)}$, approximately given by \eq{eq:c_e=S}, is proportional to the Gauss sum
\be
{\cal S}_N(\ell)\equiv\sum\limits_m w_m \exp\left[ 2\pi \rmi m^2\frac{\ell}{N}\right].\label{floquet:peg:quad}
\ee

Hence, we deal with a Gauss sum over purely quadratic phases\cite{mack+merkel}. 

In Ref.~\cite{merkel:grundlagen:2010} we have analyzed the properties of the function ${\cal S}_N={\cal S}_N(\ell)$. In particular, we have shown that ${\cal S}_N(\ell)$ allows us to factor numbers in a rather straightforward way. Since ${\cal S}_N(\ell)$ approximates the  excitation probability amplitude $c_e^{(FL)}$ of a Floquet ladder given by \eq{eq:c_efinal}, the occupation probability $|c_e^{(FL)}|^2$ at integer values $\ell$ of the dimensionless chirp $\xi$ should yield information about the factors of an appropriately encoded number $N$. 

\begin{figure}
\begin{center}
\includegraphics[width=0.65\columnwidth]{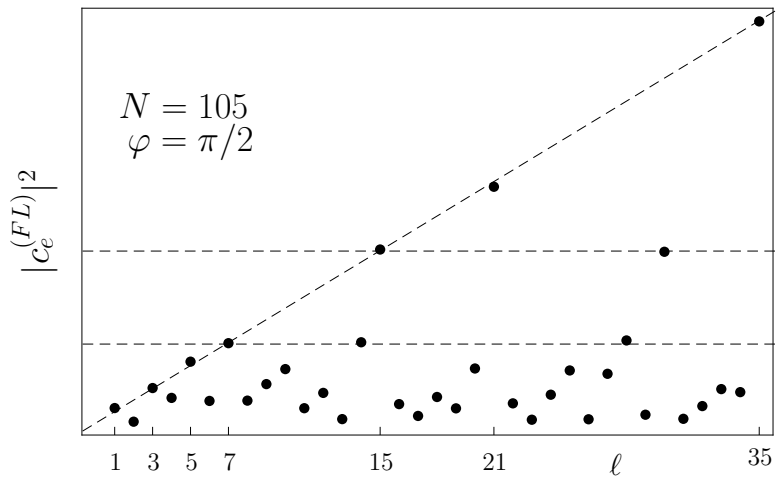}
\end{center}
\caption{Factorization of $N=105=3\cdot5\cdot7$ in the Floquet-ladder approach using the signal $|c_e^{(FL)}|^2$, \eq{eq:c_efinal}, for integer values $\xi=\ell$ of the rescaled chirp.   At the prime factors $\ell=3,\,5$ and $7$ and products $\ell=15,\,21,\,35$ the signal displays maxima. At non-factors  the signal is suppressed.   Data points corresponding to factors of $N$ are situated on a line through the origin. Integer multiples of a factor are characterized by the same value of the signal as illustrated by the two horizontal lines.. Since $c_e^{(FL)}$ is only an approximation of $S_N(\ell)$, there are small deviations of this behavior. In order to satisfy the criterion of slowly varying weight factors we have chosen the parameters $\Delta n=90$ and $\kappa=10^5\cdot 2\pi+\pi/4$.}
\label{figure105}
\end{figure}

In \fig{figure105} we verify this statement by presenting numerical results for the factorization of $N=105=3 \cdot 5 \cdot 7$ based on \eq{eq:c_efinal}. In contrast to the previous scheme the signal $|c_e^{(FL)}|^2$ is depicted only for integer values of the rescaled chirp $\xi=\ell$.
Moreover, data points with $\ell$ being a factor of $N$ and their products arrange themselves on a straight line through the origin. Data points corresponding to integer multiples of a factor are characterized by identical values. On the other hand  the signal is suppressed at non-factors of $N$ in complete accordance with the predictions of  Ref.~\cite{merkel:grundlagen:2010}.

\section{Pulse train}
\label{sec:comb}
\begin{figure}
\centerline{
\includegraphics[width=0.65\columnwidth]{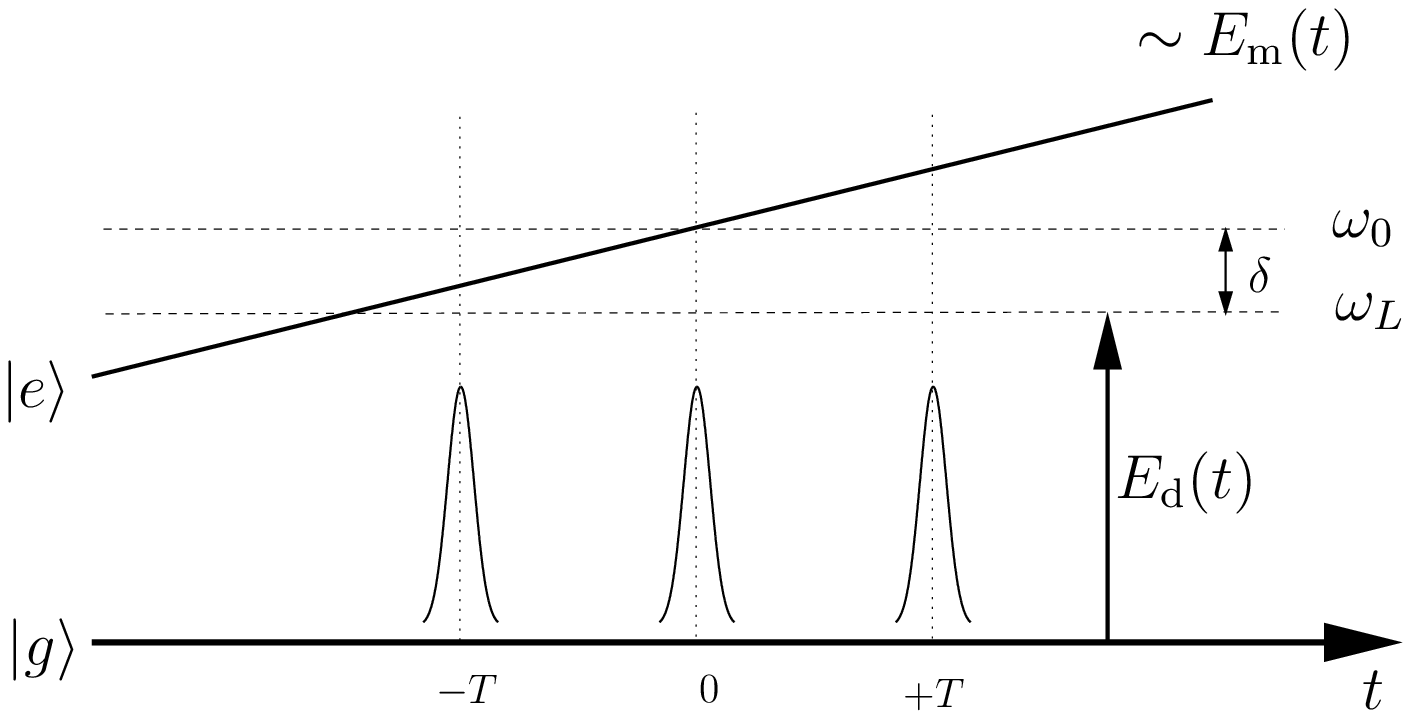}
}
\caption{Excitation by a pulse train.
We consider a two-level system with ground state $\ket{g}$ and excited state $\ket{e}$. The modulation field $E_\textrm{m}=E_\textrm{m}(t)$, \eq{driving:linear}, causes a  linear variation of the excited state energy. Simultaneously a  sequence of $2M+1$ delta-shaped weak laser pulses $E_\textrm{d}$ with envelope $h=h(t)$, \eq{envelope:pulsetrain} and carrier frequency $\omega_L=\omega_0-\delta$ induces a transfer to $\ket{e}$. Consecutive pulses here depicted for $M=1$ are separated by $T$.}
\label{figure2}
\end{figure}

In this section we turn to yet another  realization of a Gauss sum in a physical system. In contrast to the method of the preceding section now the quadratic phase factors are not due to a chirped laser pulse, but arise from the combination of  a linear time-variation of the resonance condition and a pulse train as shown by \fig{figure2}. The probability amplitude $c_e^{(PT)}$ of excitation after a sequence of laser pulses follows from the sum over the contributions from the individual pulses and is of the form of a Gauss sum. Again the system is capable of factoring numbers. However, the roles of the trial factor  and the number to be factored are interchanged.

\subsection{Excitation probability in the weak field limit}

We modulate the energy of the excited state 
by an electric field
\begin{equation}
E_\textrm{m}(t)\equiv {\cal F}_0\, \frac{t}{T},
\label{driving:linear}
\end{equation}
which increases linearly in time. Here ${\cal F}_0$ denotes the amplitude of the field and $T$ is a time scale. 

When we substitute this field into the definition \eq{beta} of the phase $\beta$ we find  the expression 
\be
\beta(t)=\frac{1}{2}\frac{\Omega_{ee}}{T} t^2-\beta_0\equiv \alpha(t)-\beta_0
\ee
which contains  the { \it time-independent} Rabi frequency  $\Omega_{ee}\equiv\wp_{ee}\,{\cal F}_0 /\hbar$ and $\beta_0 \equiv \Omega_{ee}t_0^2/(2T)$.

Moreover, we drive the one-photon transition with the electric field $E_\textrm{d}$, given by \eq{eq:Ed} and  consisting of a train 
\be
h(t)\equiv \frac{1}{2M+1}\sum\limits_{n=-M}^M \delta(t-n\,T)
\label{envelope:pulsetrain}
\ee
of $2M+1$ delta-shaped pulses separated by $T$. Here, we have chosen a normalization
\be
\int\limits_{-\infty}^{\infty}\rmd t\;h(t)  = 1
\ee
The approximation of the pulse by a delta function reflects the fact that the temporal width of the individual pulses has to be small compared to  $T$.

When we substitute the pulse train $h=h(t)$, \eq{envelope:pulsetrain}, into \eq{master:solution} and perform the integration we arrive at 
\be
c_e^{(PT)}(t)=i\,\Omega_{ge}\,\ehoch{\rmi \,\alpha(t)}\frac{1}{2M+1} \sum\limits_{n=-M}^M \exp\left[\rmi\,\left(\delta\, T\,n -\frac{\Omega_{ee}}{2} T\, n^2\right)\right].
\label{exprob0}
\ee

Here we have assumed that the range of integration in \eq{master:solution}  is large enough  to cover the whole pulse train.

Again the probability amplitude $c_e^{(PT)}$ of excitation involves the sum over quadratic phase factors and is therefore  a Gauss sum.

\subsection{Emergence of reciprocate Gauss sum}
\label{factor2}

Though in principle we could apply the same factorization scheme as in Sec.~\ref{sec:phase} we propose here a more powerful technique for factorization. Indeed, by a proper choice of parameters we eliminate in \eq{exprob0} the phase   linear in $n$.
For this purpose we relate the detuning $\delta$ and the pulse separation $T$
to the number $N$ to be factored by
\be
N\equiv \frac{\delta T}{2\pi}.
\ee
With this choice we find for the quadratic phase 
\be
\frac{1}{2} \Omega_{ee} T n^2=2\pi n^2 \frac{N}{\xi}
\ee
where we have introduced the dimensionless variable
\be
\xi \equiv \frac{2\delta}{\Omega_{ee}}.
\ee 
As a consequence,  the probability amplitude $c_e^{(PT)}$  for the pulse train given by \eq{exprob0} reduces to
\be
c_e^{(PT)}(t) =\rmi\, \Omega_{ge}\,\ehoch{\rmi \alpha(t)}
{\cal A}_N(\xi)
\label{popampc}
\ee
and is governed by the Gauss sum
\be
{\cal A}_{N}(\xi)\equiv \frac{1}{2M+1}
\sum\limits_{n=-M}^M \exp{\left[-2\pi \rmi\,n^2\,\frac{N}{\xi}\right]}.
\label{calA}
\ee
In contrast to the Gau{ss} sums of the preceding sections the roles of $N$ and $\xi$ are interchanged. Indeed, now the variable  $\xi$ appears in the denominator and the number $N$ to be factored in the numerator.

\subsection{Factorization}

In Sec.\ref{factor1} we have shown that the Gauss sum arising in the excitation of the Floquet ladder reveals the factors of  $N$  for a continuous tuning of the chirp $\xi$ as well as for integer values $\xi=\ell$. Likewise, the Gauss sum  ${\cal A}_N={\cal A}_N(\xi)$ defined by  \eq{calA} provides us with the factors for continuous as well as integer values of $\xi$. However, the analysis for continuous $\xi$ is more complicated and has been presented in Ref.\cite{woelk:2010}. For this reason we focus in the present section only on the discrete case.

Since the Rabi frequency $\Omega_{ee}$ is a free parameter we can adjust $\xi$ to be an integer $\ell$. As a result we arrive at the sum\footnote{ In Ref.~\cite{merkel:grundlagen:2010} we have shown that the  Gauss reciprocity relation establishes the connection between the two types of Gauss sums ${\cal S}_N$ and ${\cal A}_{N}$ of \eq{floquet:peg:quad} and \eq{calA(l)}, respectively.}
\be
{\cal A}_{N}(\ell)\equiv \frac{1}{2M+1}
\sum\limits_{n=-M}^M \exp{\left[-2\pi \rmi\,n^2\,\frac{N}{\ell}\right]}.
\label{calA(l)}
\ee

We now demonstrate that ${\cal A}_{N}$ is even more suited to factor numbers than the two Gauss sums  ${\cal S}$ or $S_N$ given by \eq{peg} and \eq{floquet:peg:quad}, respectively. Whenever the integer argument $\ell$ is a factor  $q$ of $N$ the phase of each phase factor of ${\cal A}_N$ is an integer multiple of $2\pi$. As a consequence, each term in the sum is unity. Since the sum contains $2M+1$ terms the signal at a factor $q$ of $N$ takes on the maximum value of
\be
|{\cal A}_{N}(q)|=1.
\ee

\begin{figure}
\centerline{
\includegraphics[width=0.65\columnwidth]{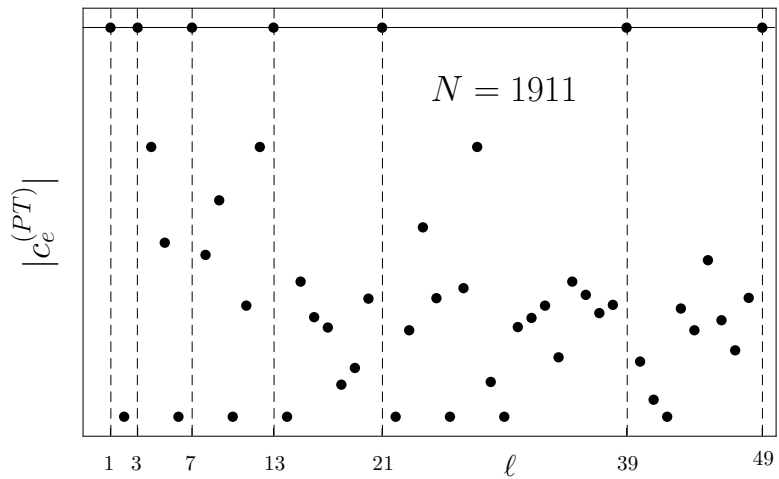}
}
\caption{Factorization of $N=1911=3 \cdot 7^2 \cdot 13$ using a  train of $21$ pulses.
The modulus $|c_e^{(PT)}|$ of the signal, \eq{popampc}, exhibits clear maxima indicated by dashed lines at integer arguments $\ell$ corresponding to factors of $N$.
}
\label{figure1911}
\end{figure}

In \fig{figure1911} we illustrate the power of this read-out mechanism of factors  using the example $N=1911=3\cdot 7^2\cdot 13$. We find that already 21   pulses allow us to decide whether $\ell$ is a factor of $N$ or not.

\section{Comparison of factorization schemes}
\label{comparison}
We devote this section to a brief comparison of the factorization schemes based on the Floquet ladder and  the pulse train discussed in Secs.\ref{sec:phase} and \ref{sec:comb}. Here we first concentrate on the methods of readout and then briefly address experimental requirements and the necessary resources.

In the Floquet-ladder approach two techniques to analyze the fluorescence signal determined by the population $|c_e^{(FL)}|^2$ in the excited state and given by \eq{eq:c_efinal}, offer themselves: (i) We measure $|c_e^{(FL)}|^2$ as a function of the continuous  chirp $\xi$. In this case pronounced maxima at trial factors indicate factors of $N$. 
(ii) An alternative read-out relies on the measurement of the signal at integer values $\xi=\ell$. Here we find that the signals at factors of $N$ form a straight line through the origin.

For the pulse-train approach the proposed read-out scheme  is based on a measurement of the signal $|c_e^{(PT)}|^2$, \eq{popampc}, at integer values of the argument $\ell$. Factors of $N$ are characterized by the same maximal value, whereas the signal at non-factors is suppressed.

Next we address the experimental requirements for these schemes to work. To reveal the factors of a given number $N$ it is sufficient to analyze the fluorescence signal  for values of the dimensionless chirp $\xi$ in the interval $[0,\sqrt{N}]$.
For the continuous version the resolution in $\xi$ has to be sufficiently high to resolve the  shape of the signal in the vicinity of candidate primes.
For the discrete scheme  the signal  has to be acquired only for integer arguments $\ell$. Nevertheless, we  require precise control of $\xi$. When we compare the number of measurements necessary in both schemes to obtain enough information for a decision on the factors, the discrete factorization schemes are favorable since less data points are required. 

It is also interesting to compare the number of terms in the Gauss sums ${\cal S}_N$ and ${\cal A}_N$ necessary to find factors. In the approach based on the Floquet ladder this number is determined by the width $\Delta n$ of the weight factor distribution  $w_m$, \eq{weight}. Indeed, this distribution has to be  sufficiently broad in order to achieve a  signal with an appropriate contrast. For the pulse-train approach the number of terms contributing to ${\cal A}_{N}$  is determined by the number of pulses in the train. Already with a few terms the signal has enough contrast to bring out the factors.

One may wonder whether the elimination of the phase linear in $n$ in the pulse-train approach  is also possible in the Floquet-ladder system. The basic idea was to chose the number $N$ to be factored such that the term linear in the summation index drops out of  the phase factor. Here, we had  three parameters at our disposal: two  are required to encode the number $N$ and one parameter is free to vary the argument $\ell$.

In the Floquet-ladder approach we also have control over the three parameters $\delta,\,\Delta\,\textrm{and}\, \phi'' $ for encoding both the number to be factored $N$ and the dimensionless argument $\ell$. However, if all three  would have been used to encode $N$, we would not have a parameter left for controlling  $\ell$.

\section{Conclusions}
\label{conclusions}

In the present article we have proposed three physical systems to implement three types of Gauss sums. Our ultimate goal was to construct an analogue computer which would calculate these Gauss sums. We have then analyzed the signal to deduce from it the factors of an appropriately encoded integer $N$.

Our first system is based on a two-photon transition in a ladder system driven by a chirped laser pulse.  Though this factorization scheme performs well for small numbers its performance for larger numbers is questionable since the required dimension $D$ of the harmonic ladder needs to be of the order of $N$. Moreover, it is rather difficult to find such an equidistant ladder system in nature.

For this reason we have investigated two other systems. In the approach of the Floquet ladder a cw-field modulates the excited state of a two-level atom giving rise to a manifold of equidistant sidebands. When driven by a chirped laser pulse the resulting excitation probability amplitude is a Gauss sum.
The second technique is based on a linear chirp of the excited state energy. A pulse train of delta-shaped pulses ensures that the excitation probability amplitude is of the form of a Gauss sum.
The origin of the quadratic phase factors is different in these two realizations of Gauss sums. In the first one they are due to the chirped laser pulse, whereas in the ladder approach they originate from a linear chirp of the resonance condition.

In all three examples the excited state probability is experimentally accessible via a detection of the fluorescence signal. Moreover, for each system we  have developed rules for determining the  factors of an appropriately encoded number.

Our factorization scheme  rests solely on interference which implies that the required resources grow exponentially with the number of digits of $N$. This feature is in contrast to Shor's algorithm which achieves an exponential speed-up due to entanglement. The next challenge is to combine these ideas with entanglement and create a Shor-algorithm with Gauss sums. However, this task goes beyond the scope of the present article and has to await a future publication.

\ack{
We have profited from numerous and fruitful discussions with M Arndt, W B Case, B Chatel, C Feiler, M Gilowski, D Haase , M Yu Ivanov, E Lutz, H Maier, M Mehring, A A Rangelov, E M Rasel, M Sadgrove, Y Shih, M $\check{\textrm{S}}$tefa\'n$\check{\textrm{a}}$k, D Suter, V Tamma, S Weber,   and M S Zubairy.
W~M. would like to thank A~Wolf for stimulating discussions.
W~M. and W~P~S acknowledge financial support by 
the  Baden-W\"{u}rttemberg Stiftung. Moreover, W~P~S also would 
like to thank the Alexander von Humboldt Stiftung and the 
Max-Planck-Gesellschaft for receiving the Max-Planck-Forschungspreis. 
I~Sh~A thanks the Israel Science Foundation for supporting this work.
Our research has also benefited immensely from the stimulating atmosphere of the \textit{Ulm Graduate School Mathematical Analysis of Evolution, Information and Complexity} under the leadership of W Arendt. 
}



\section*{References}

\bibliographystyle{apsrev}


\end{document}